\documentclass[aps,prb,amssymb,showpacs,superscriptaddress,preprint]{revtex4}
\usepackage{graphicx}
\begin{document}

\title{Including nonlocality in exchange-correlation kernel
from time-dependent current density functional theory:
Application to the stopping power of electron liquids}

\author{V.~U.~Nazarov}

\affiliation{Research Center for Applied Sciences, Academia Sinica,
 Taipei 115, Taiwan}

\author{J.~M.~Pitarke}
\affiliation {CIC nanoGUNE Consolider, Mikeletegi Pasealekua 56, E-2009 Donostia,
Basque Country}
\affiliation {Materia Kondentsatuaren Fisika Saila, UPV/EHU and Unidad Fisica de
Materiales, CSIC-UPV/EHU, 644 Posta kutxatila, E-48080 Bilbo, Basque Country}

\author{Y.~Takada}
\affiliation {Institute for Solid State Physics,
University of Tokyo, Kashiwa, Chiba 277-8581, Japan}

\author{G.~Vignale}
\affiliation {Department of Physics and Astronomy, University of Missouri,
Columbia, Missouri 65211, USA}

\author{Y.-C.~Chang}
\affiliation{Research Center for Applied Sciences, Academia Sinica,
 Taipei 115, Taiwan}

\begin{abstract}
We develop a scheme for building the scalar exchange-correlation (xc) kernel
of time-dependent density functional theory (TDDFT) from the tensorial kernel
of time-dependent {\em current} density functional theory (TDCDFT) and the
Kohn-Sham current density response function. Resorting to the local
approximation to the kernel of TDCDFT results in a nonlocal approximation to
the kernel of TDDFT, which is free of the contradictions that plague the
standard local density approximation (LDA) to TDDFT. As an application of this
general scheme, we calculate the dynamical xc contribution to the stopping
power of electron liquids for slow ions to find that our results are in
considerably better agreement with experiment than those obtained using TDDFT
in the conventional LDA.
\end{abstract}

\maketitle

\section{Introduction}

Starting with the pioneering work of Runge and Gross,~\cite{Runge-84}
time-dependent density functional theory (TDDFT) has evolved into a powerful
tool for studying excitations in atomic, molecular, and condensed-matter
systems.\cite{Petersilka-96,Wasserman-03,Reining-02,Marini-03,Burke-05}
In the linear-response regime, the key quantity of TDDFT is the dynamical
exchange and correlation (xc) kernel $f_{xc}({\bf r},{\bf r}',\omega)$
defined as the Fourier transform with respect to time of the functional
derivative
$$f_{xc}({\bf r},{\bf r}',t-t')=\frac{\delta V_{xc}({\bf r},t)}
{\delta n({\bf r}',t')},$$
where $V_{xc}$ and $n$ are the time-dependent
xc potential and particle density, respectively. In contrast with the xc
potential in static DFT, \cite{Hohenberg-64,Kohn-65} the dynamical xc potential
is strongly nonlocal with respect to space coordinates,~\cite{Perdew-84}
to the point that a local-density approximation (LDA), understood as the zeroth
order term in a regular gradient expansion, does not exist. Indeed, the use of
LDA to treat genuinely dynamical effects (i.e., effects not captured by the
adiabatic approximation) is known to lead to severe contradictions within the
theory.\cite{Vignale-95} Despite impressive successes of TDDFT, there is still
the want of a scheme for including nonlocality in xc kernels, accurate enough
and practically convenient in applications.\cite{Burke-05}

Contrary to ordinary TDDFT, the time-dependent current density functional
theory \cite{Vignale-96} (TDCDFT) is known to allow a consistent LDA, which is
believed to be of about the same level of accuracy for the time-dependent
phenomena as the standard LDA is for ground-state properties. In many concrete
applications, however, (including the calculation of the stopping power of
electron liquids described below) it is the {\em scalar} xc kernel of the
ordinary TDDFT, rather than the {\em tensorial} xc kernel of the TDCDFT, that
naturally enters the equations describing the many-body effects.

In this paper we exploit the fact that TDDFT and TDCDFT would
be completely equivalent if the exact xc functionals were known, to construct
a nonlocal approximation for the scalar xc kernel of TDDFT starting from the
LDA for the tensorial xc kernel of TDCDFT. As we shall show below, the
resulting nonlocal
xc kernel of TDDFT satisfies the
exact zero-force sum-rule, the violation of which within LDA to TDDFT had
once provided the motivation for introducing TDCDFT.\cite{Vignale-96}

We believe that our new nonlocal xc kernel has a broad range of potential
applications, particularly in transport theory.   As a first demonstration of its usefulness, we present here the
results of calculations of the stopping power  of an electron liquid
for slow ions, wherein we find that the contribution of the many-body dynamical
xc effects is not only numerically important, but also leads to better agreement
with experiment when the new nonlocal expression for $f_{xc}$ is used in lieu
of the conventional LDA.

The organization of this paper is as follows.
In Sec~\ref{SFT} we derive a formula expressing the exact scalar xc kernel of TDDFT
through the exact tensorial xc kernel of TDCDFT and the Kohn-Sham current density response function.
In Sec.~\ref{TDDFTSP} we summarize the formal TDDFT of the stopping power
of an electron liquid for slow ions and discuss the difficulties the LDA runs into.
In Sec.~\ref{disc} we give the details of our calculational procedure,
present results and their discussion. Sec.~\ref{conc} contains our conclusions.
Appendix is devoted to the interrelations between scalar
and tensorial zero-force sum rules within the exact
and approximate theories.

\section{Scalar xc kernel of TDDFT from
the tensorial xc kernel of TDCDFT}
\label{SFT}

We start from the expression of the xc kernel of TDDFT \cite{Gross-85}
(the dependence on ${\bf r}$, ${\bf r}'$, and $\omega$ is implied)
\begin{eqnarray}
f_{xc}= \chi^{-1}_{KS} - \chi^{-1} -
\frac{1}{|{\bf r}-{\bf r}'|}, \label{fsd}
\end{eqnarray}
where $\chi$ is the longitudinal density response function and $\chi_{KS}$
is its single-particle Kohn-Sham (KS) counterpart. Similarly, \cite{Vignale-96}
\begin{eqnarray}
\hat{f}_{xc,ij}=\hat{\chi}^{-1}_{KS,ij}-\hat{\chi}^{-1}_{ij}-
\frac{c}{\omega^2} \nabla_i \frac{1}{|{\bf r} - {\bf r}'|}\, \nabla'_j,
\label{ftd}
\end{eqnarray}
where $\hat{f}_{xc,ij}$ is the tensorial xc kernel of the TDCDFT,
$\hat{\chi}_{ij}$ and $\hat{\chi}_{KS,ij}$ are the many-body current density
response function and its single-particle KS counterpart, respectively.
Inverting the relation between the tensorial current density and the scalar  density response functions
\begin{eqnarray}
\chi =  -\frac{c}{\omega^2} \nabla_i \cdot \hat{\chi}_{ij}
\cdot  \nabla'_j,
\label{chit}
\end{eqnarray}
we can write
\begin{eqnarray*}
\chi^{-1} = -\frac{\omega^2}{c}
\nabla^{-2} \nabla \cdot \left(\hat{L} \hat{\chi} \hat{L} \right)^{-1} \cdot
\nabla  \nabla^{-2},
%\label{chi1}
\end{eqnarray*}
where $\hat{L}$ is the longitudinal projector operator
$\hat{L}_{ij}=\nabla_i \nabla_j \nabla^{-2}$.
Using a simple operator identity
\begin{eqnarray*}
\left(\hat{L} \hat{\chi} \hat{L} \right)^{-1} =
\hat{L} \hat{\chi}^{-1} \hat{L} -\hat{L} \hat{\chi}^{-1}
\left(\hat{T}\hat{\chi}^{-1}\hat{T}\right)^{-1}
\hat{\chi}^{-1}\hat{L},
%\label{oid}
\end{eqnarray*}
where $\hat{T}=\hat{1}-\hat{L}$ is the transverse projector,
we can write for the inverse scalar response function
\begin{eqnarray}
\chi^{-1}\!=\!
-\frac{\omega^2}{c}
\nabla^{-2} \nabla \! \cdot \!
\left[
\hat{\chi}^{-1} \!\! - \!
\hat{\chi}^{-1} \! \left(\hat{T} \hat{\chi}^{-1}\hat{T}\right)^{\!-\!1} \!\!\!
\hat{\chi}^{-1}
\right]
\! \cdot \! \nabla \nabla^{-2}
\label{chi1ts}
\end{eqnarray}
and similarly for $\chi_{KS}$.
Using Eqs.~(\ref{fsd}), (\ref{ftd}), (\ref{chi1ts}),
and the KS counterpart of the latter, we readily arrive at
\begin{eqnarray}
&&\!\!\! f_{xc} \! \! = \!
-\frac{\omega^2}{c}
\nabla^{-2} \nabla \! \cdot \!
\left\{ \!
\hat{f}_{xc}
\! + \! \left(\hat{\chi}^{-1}_{KS} \!-\!\! \hat{f}_{xc}\right)\!
\left[\! \hat{T} \! \left(\hat{\chi}^{-1}_{KS} \!-\!\! \hat{f}_{xc}
\! \right)\! \hat{T} \right]^{-1}\right. \cr\cr
&&\left. \times
\left(\hat{\chi}^{-1}_{KS} \!-\! \! \hat{f}_{xc}\right)\!-\! \hat{\chi}^{-1}_{KS}
\left(\hat{T}\hat{\chi}^{-1}_{KS}\hat{T}\right)^{-1}\!\!\!\!
\hat{\chi}^{-1}_{KS}
\right\} \cdot  \nabla \nabla^{-2}.
\label{via}
\end{eqnarray}
Equation~(\ref{via}) is
our desired and central result: it expresses the
scalar xc kernel of TDDFT in terms of its tensorial counterpart of TDCDFT.

In the case of
a {\em bounded} system, the exact scalar xc  kernel  satisfies the zero-force
sum-rule \cite{Vignale-95}
\begin{equation}
\int f_{xc}({\bf r},{\bf r}',\omega) \, \nabla' n_0({\bf r}') \, d {\bf r}'
= \nabla V_{xc}({\bf r}),
\label{srV}
\end{equation}
where $n_0({\bf r})$ and $V_{xc}({\bf r})$ are, respectively, the ground-state
density and the xc potential. On the other hand,
the tensorial $\hat{f}_{xc}$ and $\hat{\chi}_{KS}$
satisfy the corresponding sum-rules of TDCDFT.\cite{Vignale-B}
In Appendix \ref{ZFSR} we prove an important result
that with  any  {\em approximation} to the tensorial $\hat{f}_{xc}$
satisfying the zero-force sum rule, the corresponding scalar
xc kernel of Eq.~(\ref{via}) satisfies the sum-rule (\ref{srV}).

\section{TDDFT of the stopping power of electron liquid}
\label{TDDFTSP}

We now illustrate the usefulness of Eq.~(\ref{via}) by applying it to the
problem of the stopping power of an electron liquid for slow ions.

\subsection{Formal TDDFT of the stopping power of electron liquid for a slow ion}

The stopping power $dE /dx$ is the loss of energy per unit path of an ion moving through the electron liquid.
The constant of proportionality between $dE /dx$ and the ion velocity (for low velocity) defines the {\it friction coefficient} $Q$,
which can be written  as \cite{Nazarov-05}
\begin{eqnarray*}
Q  =   Q_1  +  Q_2
\end{eqnarray*}
where $Q_1$ and $Q_2$ are the single-particle and the many-body dynamical xc
contributions,  respectively.  The single-particle (binary-collisions) contribution $Q_1$ can be expressed
as \cite{Finneman-68,Ferrell-77,Echenique-81,Echenique-86}
\begin{eqnarray}
Q_1&=& \bar n_0  \, k_F \sigma_{tr}(k_F),
\label{Q1}
\end{eqnarray}
where $k_F$ is the Fermi wave-number, $\sigma_{tr}(k_F)$ is the transport
cross-section of the elastic scattering in the KS potential of an electron at
the Fermi level, and  $\bar n_0$ is the electron liquid density in the absence of the ion. As shown
in Ref.~\onlinecite{Nazarov-05}, keeping the $Q_1$ part of the friction coefficient only is equivalent
to using the {\em adiabatic} version of TDDFT.

The many-body dynamical xc contribution $Q_2$ is given by \cite{Nazarov-05}
\begin{eqnarray}
Q_2 &=& - \int [\nabla_{\bf r} n_0({\bf r})\cdot {\bf \hat v}]
[\nabla_{{\bf r}'} n_0({\bf r}')\cdot {\bf \hat v}]\cr\cr
&\times& \frac{\partial {\rm Im} f_{xc}({\bf r},{\bf r}',\omega)}
{\partial\omega} \Bigr |_{\omega=0} d{\bf r} \, d{\bf r}',
\label{Q2}
\end{eqnarray}
where $f_{xc}({\bf r},{\bf r}',\omega)$ is the scalar xc kernel of the inhomogeneous
many-body system of an ion at rest in electron liquid and ${\bf \hat v}$ is the unit vector in the direction of the ion velocity.

In the following, we focus on the calculation of $Q_2$.

\subsection{Contradiction inherent in the LDA}
The simplest approximation,
namely the  LDA to the ordinary TDDFT,\cite{Gross-85} amounts to setting
\begin{eqnarray}\label{fxc0}
f_{xc}({\bf r},{\bf r}',\omega)=f^h_{xc,L}[n_0(r),\omega]\,\delta({\bf
r}-{\bf r}'),
\end{eqnarray}
where $f^h_{xc,L}(n,\omega)$ is the $q\to 0$ limit of the longitudinal xc
kernel of a homogeneous electron liquid of density $n$. By spherical symmetry,
substitution of Eq.~(\ref{fxc0}) into Eq.~(\ref{Q2}) yields
\begin{equation}
Q_2=-\frac{4\pi}{3}  \int\limits_0^\infty dr\left[r\,n_0'(r)\right]^2
\frac{\partial {\rm Im}f^h_{xc,L}[n_0(r),\omega]}
{\partial\omega}\Bigr |_{\omega=0}.
\label{fxcl}
\end{equation}
In the limit of zero density of the electron liquid  $\bar n_0\rightarrow 0$, the
independent-electron part $Q_1$ of Eq.~(\ref{Q1}) vanishes, but $Q_2$ of
Eq.~(\ref{fxcl}) gives a finite value, because the gradient of
the ground-state density $n_0(r)$ of an isolated atom is not zero and
$\partial {\rm Im}f^h_{xc,L}(n,\omega)/\partial \omega |_{\omega=0}$ is
negative.\cite{Qian-02} Thus, LDA to the scalar $f_{xc}$ yields a finite
friction coefficient  even in the absence of the electron gas, indicating an obvious flaw of the approximation.
Table \ref{T} shows this error quantitatively for a number of atoms
in comparison with friction coefficient at $r_s=2.2$.
\begin{table}[h]
  \centering
\begin{tabular}{c c c c c c c c}
  \hline
    Atom  & He & Be & C & O & Ne & Mg & Si \\
  \hline
  Q($r_s =  \infty$) & 0.04 & 0.11  & 0.17 & 0.24 & 0.30 & 0.36 & 0.43\\
  Q($r_s$=2.2) & 0.34 & 0.43  & 0.70 & 0.46 & 0.16 & 0.15 & 0.54 \\
  \%             & 12 & 25 & 24 & 52 & 188 & 240 & 80 \\
    \hline
\end{tabular}
  \caption{\label{T}
  Inaccuracy of LDA to  TDDFT: Friction coefficient of free space
  ($r_s =  \infty$) and that of an electron liquid of $r_s  =  2.2$  for several atoms.
  Line 3 is the ratio of lines 1 and 2 (\%).}
  \label{tab}
\end{table}

To check that Eq. (\ref{via}) resolves this problem of the finite friction coefficient of
free space, it is sufficient to notice that an isolated atom is a bounded
system and hence the sum-rule (\ref{srV}) holds, which, substituted into
Eq.~(\ref{Q2}), yields zero identically. Moreover, LDA to TDCDFT satisfies
the zero-force sum-rule by construction, \cite{Vignale-96} ensuring, as is shown in the Appendix,
that Eq. (\ref{srV}) holds for a bounded system even if the local version of
TDCDFT is used in Eq.~(\ref{via}).
\footnote{Notice that
the sum-rule  (\ref{srV}) {\em does not hold} for an {\em extended} system: this is why $Q_2$
is different from zero  for an electron liquid of finite density.}

\section{Calculational procedure, results, and discussion}
\label{disc}
Our numerical procedure is to evaluate and invert operators entering
Eq.~(\ref{via}) on an ortho-normal set of radial basis functions. For
$\hat{\chi}_{KS,ij}$, we have employed the standard method of using
the static KS orbitals to build the independent-electron response function
\begin{eqnarray}
&&\hat{\chi}_{KS,ij}({\bf r},{\bf r}',\omega) =
\frac{1}{ c} \, n_0({\bf r}) \delta({\bf r}-{\bf r}') \, \delta_{ij} -\frac{1}{4 c} \times \cr\cr
&&\sum\limits_{\alpha \beta} \frac{f_\alpha-f_\beta}{\omega-\epsilon_\beta+\epsilon_\alpha+i\eta}
\left[\psi^*_\alpha({\bf r}) \nabla_i \psi_\beta({\bf r})- \psi_\beta({\bf r}) \nabla_i \psi^*_\alpha({\bf r})\right]\cr\cr
&&\times \left[ \psi^*_\beta({\bf r}') \nabla'_j \psi_\alpha({\bf r}')- \psi_\alpha({\bf r}') \nabla'_j \psi^*_\beta({\bf r}') \right],
\label{chiKS}
\end{eqnarray}
where $\psi_\alpha({\bf r})$ and $\epsilon_\alpha$ are the single-particle wave-function
and eigenenergy, respectively, in the state $\alpha$, and $f_\alpha$ is the occupation number of this state.
For $\hat{f}_{xc,ij}$, we use LDA to TDCDFT
as \cite{Vignale-97}
\begin{eqnarray}
&&\int
f_{xc,i k}({\bf r},{\bf r}',\omega)\,
j_k({\bf r}') d {\bf r}' = \frac{i c}{\omega} \times \cr\cr
&&\left[ - \nabla_i V_{xc}^{ALDA}({\bf r},\omega)+
\frac{1}{n_0({\bf r})} \nabla_k \, \sigma_{xc,ik}({\bf r},\omega) \right],
\label{G1}
\end{eqnarray}
where
\begin{eqnarray}
V_{xc}^{ALDA}({\bf r},\omega)=\frac{1}{i \omega} \,
\epsilon''_{xc}[n_0({\bf r})] \, \nabla_k \, j_k({\bf r}),
\end{eqnarray}
$\epsilon_{xc}(n)$ is the xc energy density,
\begin{eqnarray}
&&\sigma_{xc,i k}({\bf r},\omega)=
\tilde{\eta}_{xc}[n_0({\bf r}),\omega] \left[\nabla_k\, u_i({\bf r})
+ \nabla_i\, u_k({\bf r})\right.\cr\cr
&&\left. -\frac{2}{3} \, \nabla_s u_s({\bf r}) \, \delta_{i k} \right]+
\tilde{\zeta}_{xc}[n_0({\bf r}),\omega] \, \nabla_s u_s({\bf r}) \, \delta_{i k},
\label{sig}
\end{eqnarray}
is the stress tensor, and ${\bf u}({\bf r}) ={\bf j}({\bf r})/n_0({\bf r})$ is the velocity field.
The viscosity coefficients are given by
\footnote{ We have used the zero-temperature
viscosities, which are valid for $\omega\gg \! T^2/E_F$.
This is justified since in the experimentally accessible regime
$\omega \sim 0.1$ a.u. and $T^2/E_F \sim 10^{-6}$ a. u. at room temperature.}
\begin{eqnarray}
&&\tilde{\zeta}_{xc}(n,\omega)=-\frac{n^2}{i\omega}
\left[f_{xc,L}^h(n,\omega) -\frac{4}{3}f_{xc,T}^h(n,\omega) -
\epsilon''_{xc}(n) \right],\cr\cr
&&\tilde{\eta}_{xc}(n,\omega)=-\frac{n^2}{i\omega} f_{xc,T}^h(n,\omega),
\label{zeet0}
\end{eqnarray}
and $f_{xc,T}^h(n,\omega)$ is the transverse xc kernel of the homogeneous electron liquid
of density $n$.

\begin{figure}[h]
\includegraphics[width=8 cm,,trim=20 10 10 0]{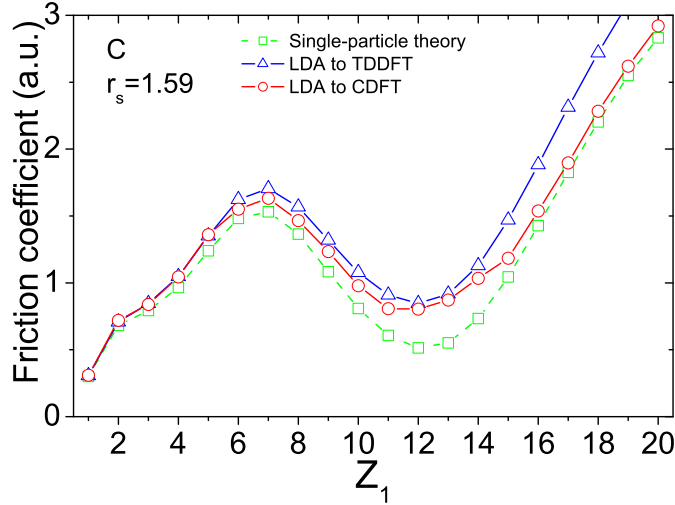}
\caption{\label{Fig_C} (Color online) Friction coefficient of electron liquid of carbon density
($r_s=1.59$) versus the atomic number of an ion. Green squares are results with
neglect of the dynamical xc [Eq.~(\ref{Q1})]. Blue triangles are results with the
dynamical xc included within  LDA to the conventional TDDFT
[Eq.~(\ref{fxcl})]. Red circles are results with the dynamical xc included with
use of Eqs. (\ref{via}) and (\ref{G1})-(\ref{zeet0}).}
\end{figure}

In Fig.~\ref{Fig_C}, we plot the results at $r_s=1.59$ corresponding to the
valence electron-density of carbon.\footnote{The available experimental stopping power of
carbon
[D. Ward {\it et al.}, Can. J. Phys. {\bf 57}, 645 (1979);
G. H{\"{o}}gberg, Phys. Status Solidi B {\bf 46}, 829 (1971)]
is predominantly determined by collisions
with lattice atoms, making it meaningless to compare with electron gas model
calculations.} It is instructive that within $1 \le Z_1 \le 14$ both
TDCDFT and TDDFT give virtually the same result, which, we believe, is
generally true for light atoms in high-density electron liquid. Then, at higher $Z_1$,
rather abruptly, the dynamical xc contribution almost vanishes in our present
calculation, which can be understood qualitatively recalling that for heavy
atoms in the electron liquid the charge-density distribution is close to that of isolated atoms,
and hence $Q_2$ should be small.

\begin{figure}[h]
\includegraphics[width=8 cm,trim=20 10 10 0]{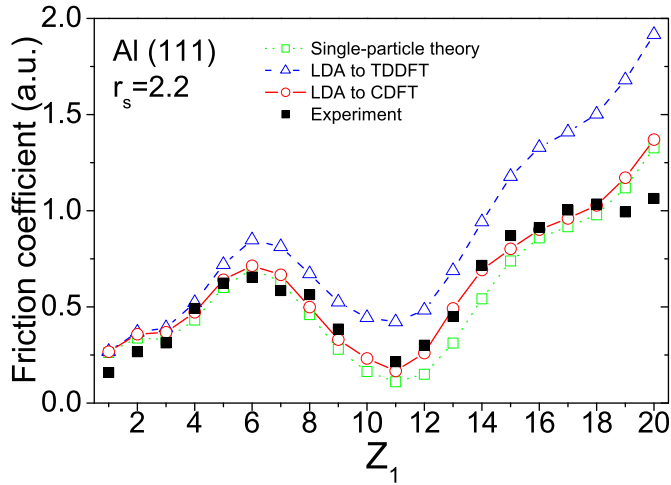}
\caption{\label{Fig_Al}
(Color online) Friction coefficient of electron liquid of $r_s=2.2$ versus the atomic number of an ion.
The green squares are the results with  neglect of the dynamical xc as obtained
from Eq.~(\ref{Q1}). Blue triangles are the results with the dynamical xc included
within LDA to the conventional TDDFT as obtained from Eq.~(\ref{fxcl}).
Red circles are results with the dynamical xc included with use of Eq. (\ref{via})
for xc kernel and Eqs. (\ref{G1})-(\ref{zeet0}) of LDA to TDCDFT. Black solid
squares are the measured stopping power of Al of Ref.~\onlinecite{Winter-03} for ions
($v=0.5~{\rm a.u.}$) moving at a distance of 1.2  a.u. from the last atomic
plane of the Al (111) surface.}
\end{figure}

In order to compare our result with experiment, we plot the friction coefficient at $r_s=2.2$ versus
the atomic number of a moving ion in Fig.~\ref{Fig_Al}. Results of the
calculations with neglect of the dynamical xc [Eq.~(\ref{Q1})], LDA to TDDFT
[the sum of Eq.~(\ref{Q1}) and Eq.~(\ref{fxcl})], and  LDA to TDCDFT
[the sum of Eq.~(\ref{Q1}) and Eq.~(\ref{Q2}) with $f_{xc}$ given by
Eqs. (\ref{via}) and (\ref{G1})-(\ref{zeet0})] are shown, together with
the experimental data of Ref. \onlinecite{Winter-03} for ions moving with
the velocity of 0.5 a.u. at the distance of 1.2 a.u. from the last atomic
plane of the (111) surface of aluminum. The inhomogeneity of the electron
density an ion travels through is weak under these conditions, and we have
used $r_s$ estimated experimentally. \cite{Winter-03} Moreover, the
experimental stopping power is predominantly electronic since the trajectory of an ion
is well separated from the lattice atoms. All together, these two conditions
justify the comparison with the theory within the electron liquid model. The non-monotonic
dependence of the friction coefficient on the atomic number of the ion (so-called $Z_1$-oscillations)
is known to result within the single-particle theory from the competition
between the increase in the electron liquid-ion interaction with the growing charge of the
nucleus of the ion and  its decrease due to the screening by shells of bounded
electrons of the pseudo-atom as well as its resonant states. \cite{Echenique-86}

While LDA to TDDFT (triangles in Fig.~\ref{Fig_Al}) largely overestimates the friction coefficient
at $Z_1\ge5$, the results using Eq. (\ref{via}) (circles in Fig.~\ref{Fig_Al})
are in good agreement with the experiment in a wide range of $3\le Z_1 \le
18$. A deviation occurs at small and large $Z_1$, where the experimental friction coefficient
is {\em lower} than the independent-electrons calculations (open squares in
Fig.~\ref{Fig_Al}). This feature has recently been reported as due to the
finite velocity of ions.\cite{Vincent-07} Hence it is an effect of the
deviation from linear dependence of stopping power on velocity. The same effect gives a
{\em positive} contribution at $8\le Z_1 \le 12$, suggesting that combined with
the many-body effects of the present theory the agreement with experiment can
be further improved.\footnote{Ref. \onlinecite{Vincent-07}
attributes the overestimation by Ref. \onlinecite{Nazarov-05} of the
role of the many-body effects to the use of the {\em total}
ground-state density rather than that of the delocalized states only. The
total density is, however, the basic variable of  TDDFT and without any
assumptions it enters the rigorous result of Eq.~(\ref{Q2}). The real source
of the overestimation of the dynamical xc in Ref. \onlinecite{Nazarov-05} was,
as Ref.~\onlinecite{Nazarov-05} had anticipated and the present work shows,
use of LDA within TDDFT.} In the range $13\le Z_1 \le 17$, the dynamical
many-body effects seem to be solely responsible for the enhancement of the friction coefficient
compared with the independent-electron theory.

\begin{figure}[h]
\includegraphics[width=8 cm,trim=20 10 10 0]{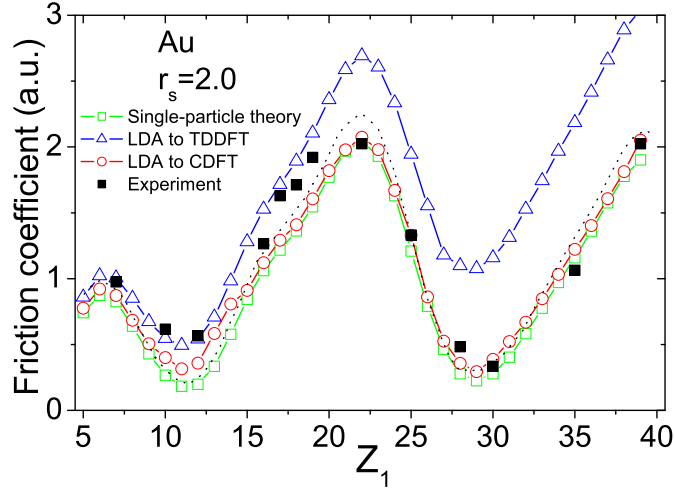}
\caption{\label{Fig_Au} (Color online) Friction coefficient of electron liquid of $r_s=2$ versus the
atomic number of an ion. Green squares are the results with neglect of the
dynamical xc [Eq.~(\ref{Q1})]. Blue triangles are the results with the dynamical
xc included within LDA to the conventional TDDFT [Eq.~(\ref{fxcl})]. Red circles
are results with the dynamical xc included with use of Eqs. (\ref{via})
and (\ref{G1})-(\ref{zeet0}). Black solid squares are the measurements from
Ref.~\onlinecite{Bottiger-69} of stopping power of Au for ions ($v=0.68~{\rm a.u.}$)
channeled along the (110) direction. The black dotted line is the calculation of
Ref.~\onlinecite{Nagy2-89} with the dynamical xc included within the
linear-response theory of the homogeneous electron gas.}
\end{figure}

In Fig.~\ref{Fig_Au}, we plot the friction coefficient at $r_s =2$ versus the atomic number
of ions in the range $5\le Z_1 \le 39$. This is compared with the available
measured stopping power for ions with the velocity of $0.68~{\rm a.u.}$ channeled along
the (110) direction in gold. Due to the channeling, collisions with the
lattice atoms again do not give significant contribution to the stopping power. It must be
noted, however, that under channeling conditions the assumption of the nearly
constant electron density is an uncontrolled approximation. One important
qualitative conclusion we can draw from comparison of the theory and
experiment in this case is that for $Z_1 \ge 22$ the role of dynamical xc
effects becomes negligible in both experiment and the present theory,
while LDA to the TDDFT yields these effects largely overestimated. Similar
to the dip in Fig.~\ref{Fig_Al}, the underestimated theoretical values at
$7\le Z_1\le 12$ can be attributed to the effect of finite
velocity.\cite{Vincent-07} However, within the range $16 \le Z_1 \le 19$ the
dynamical xc contribution is too small to account for the onset at the
experimental data, nor can the persistent enhancement of the friction coefficient in this range be
attributed to the effect of finite velocity within the independent-particle
theory. Further studies are required to elucidate the nature of this onset,
the inhomogeneity of electron density being the most plausible cause.
The dotted line in Fig.~\ref{Fig_Au} represents the friction coefficient of silver
obtained with the dynamical electron-electron interactions included in Ref.~\onlinecite{Nagy-89}
within the framework of the linear-response theory of the homogeneous electron gas.
In the case of the non-degenerate plasma, an approach similar to that of Ref.~\onlinecite{Nagy-89} has been reported in Refs.~\onlinecite{Morawetz-96} and \onlinecite{Selchow-99}.

\section{Conclusions}
\label{conc}
We have rigorously expressed the dynamical xc
kernel $f_{xc}({\bf r},{\bf r}',\omega)$ of TDDFT in the terms of its TDCDFT
tensorial counterpart and the Kohn-Sham current density response function of
independent electrons. Then, using the local density approximation to
TDCDFT, we have built a nonlocal approximation to $f_{xc}({\bf r},{\bf r}',
\omega)$ which satisfies the exact zero-force sum-rule for bounded systems.
We believe that our new  approximation will be broadly applicable to a variety of problems in electronic transport theory.

As a first application, we have calculated the dynamical xc contribution to the stopping power of an electron liquid for slow ions.  In doing so
we have resolved a basic difficulty of the conventional LDA  --  the finite friction coefficient of free space --  and we have improved  the overall agreement between theory and experiment.

\appendix
\section{Relation between scalar and tensorial zero-force sum-rules}
\label{ZFSR}
The following sum-rules hold\cite{Vignale-B} for the exact tensorial xc kernel
 \begin{eqnarray}
\int
\hat{f}_{xc,ij}({\bf r},{\bf r}',\omega) \, n_0({\bf r}') d {\bf r}'=
-\frac{c}{\omega^2} \nabla_i \nabla_j V_{xc}({\bf r})
\label{sumbrf}
\end{eqnarray}
and for the KS and the interacting current density response functions, respectively,
\begin{eqnarray}
&&\frac{c}{\omega^2} \int
\hat{\chi}_{KS,ik}({\bf r},{\bf r}',\omega) \, \nabla'_k \nabla'_j V_{KS}( {\bf r}')
\, d {\bf r}' = \nonumber \\
&&c \int
\hat{\chi}_{KS,ij}({\bf r},{\bf r}',\omega) \, d {\bf r}' - n_0({\bf r})\,\delta_{ij},
\label{sumbrKS}
\end{eqnarray}
\begin{eqnarray}
&&\frac{c}{\omega^2} \int
\hat{\chi}_{ik}({\bf r},{\bf r}',\omega) \, \nabla'_k \nabla'_j V_0( {\bf r}')
\, d {\bf r}' = \nonumber \\
&&c \int
\hat{\chi}_{ij}({\bf r},{\bf r}',\omega) \, d {\bf r}' - n_0({\bf r})\,\delta_{ij},
\label{sumbr0}
\end{eqnarray}
where $V_0({\bf r})$ is the bare potential.

In this Appendix, we prove that for any {\em approximation} to $\hat{f}_{xc}$ satisfying
the sum-rule (\ref{sumbrf}), the corresponding scalar $f_{xc}$ obtained through Eq.~(\ref{via})
satisfies the sum-rule of  Eq.~(\ref{srV}).
First, the validity of Eq.~(\ref{sumbrKS}) is independent on an approximation
for $\hat{f}_{xc}$, and it can be verified directly with use of the explicit
representation of $\hat{\chi}_{KS}$ of Eq.~(\ref{chiKS}).
Second, Eq.~(\ref{sumbr0}) holds if Eqs.~(\ref{sumbrf}) and (\ref{sumbrKS}) hold as can be  seen
by easily inverting the arguments of Ref.~\onlinecite{Vignale-B} leading from Eqs.~(\ref{sumbrKS}) and
(\ref{sumbr0}) to Eq.~(\ref{sumbrf}). Equation~(\ref{sumbr0}) can be rewritten as
\begin{eqnarray}
&&\frac{c}{\omega^2} \int
\hat{\chi}_{ik}({\bf r},{\bf r}',\omega) \, \nabla'_k \nabla'_j V_0( {\bf r}')
\, d {\bf r}' = \cr\cr
&&c \int
\hat{\chi}_{ik}({\bf r},{\bf r}',\omega) \,\nabla'_k r'_j \, d {\bf r}' - n_0({\bf r})\,\delta_{ij}.
\label{AAAA}
\end{eqnarray}
The next step involves integration by parts requiring
the response function to vanish at infinity and, therefore, it applies to bounded systems only.
In this case we can write
multiplying Eq.~(\ref{AAAA}) scalarly from the left by $\nabla$ and using  Eq.~(\ref{chit})
\begin{eqnarray}
\int
\chi({\bf r},{\bf r}',\omega) \left[ \omega^2 r'_j -\nabla'_j V_0( {\bf r}')  \right]
\, d {\bf r}' =
\nabla_j n_0({\bf r}),
\nonumber
\end{eqnarray}
and after the inversion
\begin{eqnarray}
\int
\chi^{-1}({\bf r},{\bf r}',\omega) \nabla_j n_0({\bf r}')
\, d {\bf r}' =
\omega^2 r_j -\nabla_j V_0( {\bf r}).
\label{Achi}
\end{eqnarray}
A similar relation holds for $\chi_{KS}$
\begin{eqnarray}
\int
\chi_{KS}^{-1}({\bf r},{\bf r}',\omega) \nabla_j n_0({\bf r}')
\, d {\bf r}' =
\omega^2 r_j -\nabla_j V_{KS}( {\bf r}).
\label{AchiKS}
\end{eqnarray}
Subtracting Eq.~(\ref{Achi}) from Eq.~(\ref{AchiKS}) and using the definition of
Eq.~(\ref{fsd}), we immediately arrive at Eq.~(\ref{srV}).

\acknowledgments

G. V. and Y. T. acknowledge, respectively, financial support by the Department
of Energy grant DE-FG02-05ER46203 and a Grant-in-Aid for Scientific
Research in Priority Areas (No.17064004) of MEXT, Japan.

\end{document}